\documentclass[aps,prb,twocolumn,showpacs,superscriptaddress,amsmath,amssymb]{revtex4}

\usepackage{graphicx}
\usepackage{dcolumn}
\usepackage{bm}
\usepackage{ifthen}
\usepackage{booktabs}

\begin{document}

\title{Atomistic pseudopotential calculations of the optical properties of 
InAs/InP self-assembled quantum dots}
\author{Ming Gong}
\affiliation{Key Laboratory of Quantum Information,
University of Science and Technology
of China, Hefei, 230026, People's Republic of China}
\author{Weiwei Zhang}
\affiliation{Key Laboratory of Quantum Information,
University of Science and Technology
of China, Hefei, 230026, People's Republic of China}
\author{Zhuming Han}
\affiliation{Key Laboratory of Quantum Information,
University of Science and Technology
of China, Hefei, 230026, People's Republic of China}
\author{G. C. Guo}
\affiliation{Key Laboratory of Quantum Information, 
University of Science and Technology
of China, Hefei, 230026, People's Republic of China}
\author{Lixin He\footnote{Electronic mail: helx@ustc.edu.cn.}}
\affiliation{Key Laboratory of Quantum Information, 
University of Science and Technology
of China, Hefei, 230026, People's Republic of China}
\date{\today }

\begin{abstract}
We present a comprehensive study of the optical properties of 
InAs/InP self-assembled quantum dots (QDs) using an empirical pseudopotential method and 
configuration interaction treatment of the many-particle effects. The results
are compared to those of InAs/GaAs QDs. 
The main results are:
(i) The alignment of emission lines of neutral exciton, charged exciton and biexciton
in InAs/InP QDs is quite different from that in InAs/GaAs QDs.
(ii) The hidden correlation in InAs/InP QDs is 0.7 - 0.9 meV, smaller than that
in InAs/GaAs QDs.
(iii) The radiative lifetimes of neutral exciton, charged exciton and biexciton in InAs/InP QDs 
are about twice longer than those in InAs/GaAs QDs. 
(v) The phase diagrams of few electrons and holes in InAs/InP QDs differ greatly from those in 
InAs/GaAs QDs. The filling orders of electrons and holes are shown to obey the Hund's rule 
and Aufbau principle, and therefore the photoluminescence spectra of highly charged excitons are
very different from those of InAs/GaAs QDs.

\end{abstract}

\pacs{73.21.La, 71.35.Pq, 78.67.Hc}



\maketitle


\section{Introduction}
\label{introduction}

Self-assembled quantum dots (QDs) have attracted great
interest in the past two decades due to their potential applications in
the optoelectronic devices \cite{huffaker98, djie07, shields07} and
quantum information science,\cite{bonadeo00, xu07a, xu07b, atature06},
such as high-efficiency and low-threshold lasers, \cite{huffaker98} 
single photon source,\cite{michler00, imamoglu94, yuan02} entangled photon 
emitters,\cite{stevenson06,akopian06}  and qubits etc.\cite{loss98, xu07a, xu07b} 
These applications all take the advantages of the unique properties 
of the QDs which have discrete energy levels 
and sharp absorption/emission lines\cite{finley01,ware05, gammon96a}
due to the three-dimensional confinement effects. 

Until very recently, most of the experimental and theoretical works focus on
InAs/GaAs QDs \cite{bayer02a, bayer02c, bonadeo00, huffaker98, reuter05, fonseca98, 
bock02, stier99, williamson99, wang99a, williamson00, williamson01, he04a}.
Huge progress has been made both in understanding of the fundamental
physics and towards real applications of these QDs. For examples, the
entangled photon sources from biexciton cascade process have been demonstrated
experimentally.\cite{stevenson06, akopian06} 
 The strong coupling
between QDs and cavity has been achieved. \cite{reithmaier04,yoshie04,
 hennessy07}
The coherently manipulation of
single charge and spin in InAs/GaAs QDs have also been demonstrated.
\cite{kamada01, zrenner02, stievater01,xu07b, emary07, ramsay08}
These achievements make QDs extraordinary promising
candidates for quantum information applications. 
Theoretically, the electronic structures and optical properties of InAs/GaAs QDs
have been studied intensively via both ${\bf k} \cdot {\bf p}$ 
methods, \cite{bahder90,nakaoka04,pryor98b,pryor99,pryor97} and the 
microscopic models.\cite{wang99b,williamson00,bryant03}

On the other hand, InAs/InP QDs have attracted interest 
only very recently \cite{gong04,caroff05,kim09, reimer09,cornet06} 
mainly because their emission wavelengths are naturally around 1.55 $\mu$m ($C$-band), 
which are ideal for fiber optical telecommunication applications. 
Compared to InAs/GaAs QDs, there are much fewer works on InAs/InP QDs,
both experimentally and theoretically. 
It has been shown from atomistic pseudopotential calculations that 
the electronic structures of InAs/InP QDs
differs greatly from those of InAs/GaAs QDs,\cite{gong08} even though they have
the same dot material, but different matrix material. 
Interestingly, it has been found \cite{he08} that the fine structure
splittings (FSS) in
InAs/InP(001) QDs are very small, which is very
suitable for $C$-band entangled photon source. 
Indeed, recent experiments indicate that InAs/InP QDs do not have significant FSS.\cite{sasakura09}

However, the optical properties of InAs/InP QDs has not yet been
systematically studied. 
In this work, we present a comprehensive study of the
optical properties of InAs/InP QDs
using an atomistic empirical pseudopotential method and the configuration
interaction (CI) treatment of the many-particle effects.\cite{wang99b}
The methods have been successfully applied to
InAs/GaAs QDs, and obtained many results that are very good agreement with experiments.
\cite{ediger05, ediger07a}
A nice review of the methods can be found in Ref. [\onlinecite{bester09}].
We find that the photoluminescence (PL) spectra of InAs/InP QDs are very different from those of
InAs/GaAs QDs in many aspects, including 
the alignment of emission lines of the neutral exciton, charged excitons and
biexciton, the hidden correlation, the radiative lifetimes, and 
the highly charged PL spectra.
We hope this work can provide helpful insight for the optical
properties of InAs/InP QDs.

The rest of the work is organized as follows. In Sec. \ref{theory}, we outline the
methods used in our calculations. In Sec. \ref{bindingenergy} and Sec. \ref{correlation}
we discuss the binding energies and the hidden correlation in InAs/InP QDs.
The radiative lifetimes of exciton, biexciton, and trions are discussed in
Sec. \ref{lifetime}. The phase 
diagrams of electrons and holes in InAs/InP QDs are presented in  Sec. \ref{phase-diagram},
whereas the highly charged exciton PL spectra in InAs/InP QDs are discussed in
Sec. \ref{multi-x}. 
We summarize in Sec. \ref{conclusion}.

\section{Methods}
\label{theory}

\begin{figure}
\centering
\includegraphics[width=0.4\textwidth]{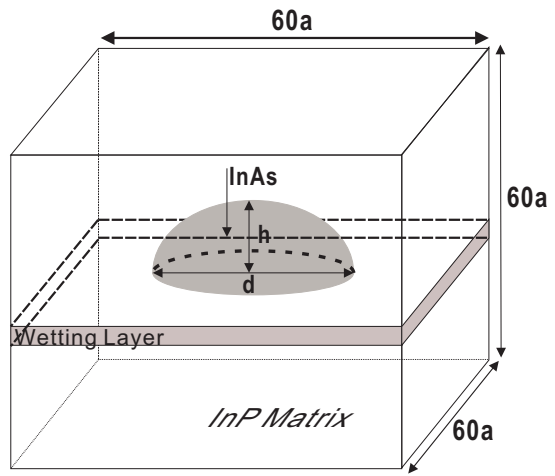}
\caption{A sketch of the geometry of a lens-shaped InAs QD embedded in the InP
matrix on the top of two-monolayer wetting layer.}
\label{fig:qd-in-matrix}
\end{figure}

Figure \ref{fig:qd-in-matrix} depict the geometry of the QD system used in
our calculations. The lens-shaped InAs QD is embedded in the 
center of the 60 $\times$60$\times$60 InP 8-atom unit cell on the top of a 
two-monolayer wetting layer. 
The single-particle electronic structures of the QDs are calculated via 
an empirical pseudopotential 
method,\cite{wang99b, williamson00} whereas the many-particle effects are treated
via a configuration interaction method, as follows.

We first obtain the atomic positions of the QDs structure. 
This is done by minimizing the strain energy of the system via
the valance force field (VFF) method,\cite{keating66, martin70}  
which has been demonstrated to be a good
approximation for semiconductors.
Once we obtain the relaxed positions ${\bf R}_{i,\alpha}$ of all the atoms
of type $\alpha$ at site $i$, we calculate the total electron-ion
potential $V_{\rm ps}({\bf
r})=V_{\rm SO}+\Sigma_{i}\Sigma_{\alpha}\upsilon_{\alpha}({\bf r}-{\bf R}_{i,\alpha})$ 
as a superposition of the local screened
atomic pseudopotentials $\upsilon_{\alpha}({\bf r})$, the total (nonlocal)
spin-orbit (SO) potential $V_{\rm SO}$. \cite{williamson00} The Schr\"{o}dinger equation,
\begin{equation}
\left[ -{1 \over 2} \nabla^2
 + V_{\rm ps}({\bf r}) \right] \psi_i({\bf r})
=\epsilon_i \;\psi_i({\bf r}) \; , \label{eq:schrodinger}
\end{equation}
is solved by using a linear combination of bulk bands (LCBB)
method, \cite{wang99b}\  in a basis
$\{\phi_{n,\tensor{\epsilon},\lambda} ({\bf k})\}$ of Bloch orbitals
of band index $n$ and wave vector ${\bf k}$ of material $\lambda$
(= InAs, InP), strained uniformly to strain $\tensor{\epsilon}$.
For each material $\lambda$, we used a basis set of $n$= 6 bands
(including spin) for the hole states and $n$= 2 for the electron
states, on a $6\times6\times16$ {\bf k}-point mesh around the $\Gamma$ point.
It has been shown that the energy levels changes in InAs/GaAs
QDs due to the piezoelectric effects are quite small. \cite{bester05a}
Because the lattice mismatch in the InAs/InP QDs is only half
of that of the InAs/GaAs QDs, 
we expect that the  piezo-effect should be even smaller in the InAs/InP dots, 
and therefore the piezoelectric potential is neglected in the calculations.

Due to the spatial confinement, the carriers in the QDs have strong Coulomb
interactions. 
The many-particle Hamiltonian reads,
\begin{equation}
H = \sum_{a, \sigma}\epsilon_{a} \psi_{a, \sigma}^{\dag}\psi_{a, \sigma} +
{1\over 2} \sum_{i,j,k,l}
\Gamma_{k, l}^{i,j} \psi_{i}^{\dag}\psi_{j}^{\dag} \psi_{k}\psi_{l},
\label{eq-ci}
\end{equation}
where $\epsilon_{a}$ is the single particle levels obtained in Eq. (\ref{eq:schrodinger}). 
$\Gamma_{k, l}^{i,j}$ is the Coulomb integral matrix, i.e.,
\begin{equation}
\Gamma_{k, l}^{i,j} = \sum_{\sigma, \sigma'}\int d {\bf r} d{\bf r'}
      {\psi_{i}^*({\bf r}, \sigma) \psi_{j}^*({\bf r'}, \sigma')  
\psi_{k}^*({\bf r}, \sigma) \psi_{l}^*({\bf r'}, \sigma')  \over
\epsilon({\bf r}-{\bf r'}) |{\bf r-r'}|},
\label{eq-gamma}
\end{equation}
where $\epsilon({\bf r}-{\bf r'})$ is the full screened dielectric constant.
\cite{franceschetti99}
The many-particle Hamiltonians are solved by using a configuration interaction (CI) 
method in which the Hamiltonians are diagonalized 
in the Slater determinant basis.\cite{franceschetti00} 
This method has been successfully applied to study the optical properties of InAs/GaAs QDs, 
and the obtained results are in excellent agreement with
experiments.\cite{ediger05,ediger07a,ediger07b,ding10,williamson00} 

\section{Results}
\label{results}

\begin{figure}
\centering
\includegraphics[width=0.48\textwidth]{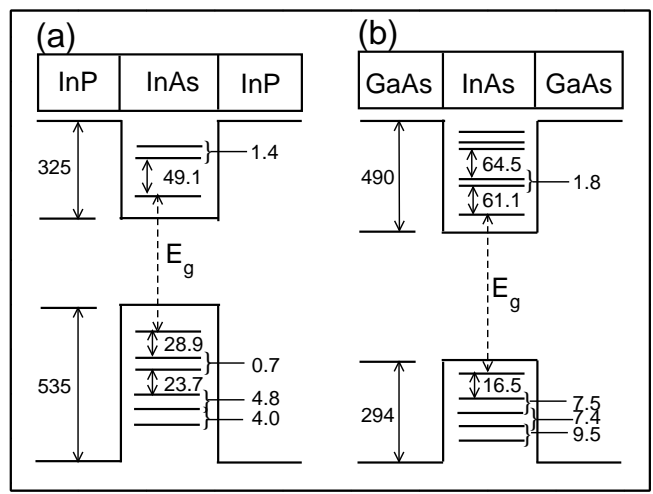}
\caption{The single-particle energy levels (in meV) in a lens-shaped (a) InAs/InP QD
  and (b) InAs/GaAs QD, with diameter $D$= 25 nm and height $h$= 2.5 nm.}
\label{fig-energy-diag}
\end{figure}

Figure \ref{fig-energy-diag} (a) and (b) show the single-particle energy
levels for typical InAs/InP and InAs/GaAs QDs, respectively, with height $h$=2.5 nm and
diameter $D$=25 nm.  The differences
between the single-particle electronic structures of the two dots have been
addressed in Ref. [\onlinecite{gong08}]. Here we merely outline the 
main differences between the two dots that are related to this work.
(i) The confinement of electron and hole in InAs/GaAs QDs is 490 and 294 meV, while in InAs/InP 
QDs is 325 and 535 meV, respectively. Therefore, holes (electrons) are much
more (less) confined in InAs/InP QDs.
(ii) Due to the much smaller confinement potential for electrons in the InAs/InP
QDs, fewer electron states are confined.
(iii) The $p$-orbit energy splitting of electrons and holes in InAs/GaAs QDs
are about 1.8 meV and 7.5 meV respectively, whereas 
in InAs/InP QDs, they are 1.4 meV and 0.7 meV, respectively. 
(iv) The hole states in tall InAs/GaAs QDs can be localized in the interface, but in 
InAs/InP QDs, hole levels have well defined shell structure and no localization
has been observed. These differences can greatly influence the optical properties of 
InAs/InP QDs, as will be shown in the following sections.

\subsection{Binding energies of $X^+$, $X^-$ and $XX$}
\label{bindingenergy}

\begin{figure}
\centering
\includegraphics[width=0.42\textwidth]{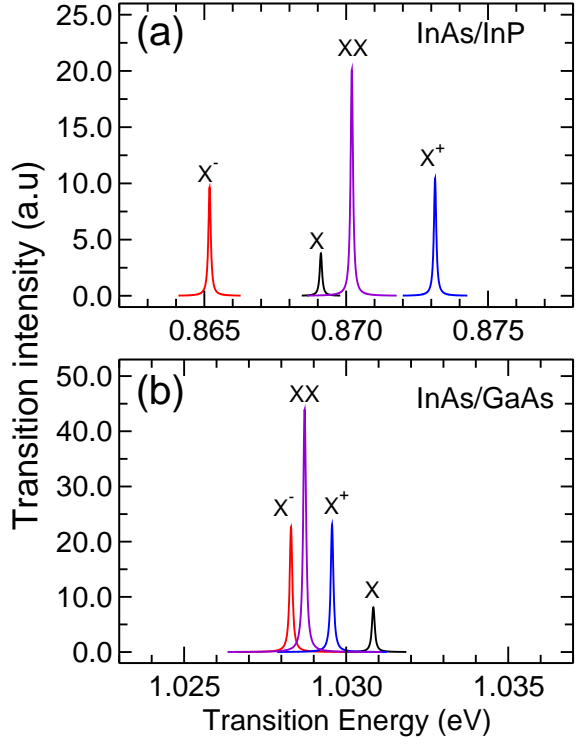}
\caption{(Color online) The PL spectra of $X$, $XX$, $X^+$ and $X^-$ in a lens-shaped 
(a) InAs/InP QD and (b) InAs/GaAs QD, with 
$D$= 25 nm and $h$ = 2.5 nm}
\label{fig-energy-pl}
\end{figure}

A typical PL spectra for lens-shaped InAs/InP and
InAs/GaAs QDs with $D = 25$ nm and $h = 2.5$ nm are presented in Fig. \ref{fig-energy-pl}
(a) and (b), respectively. We focus on the alignment of emission line of $\chi^q$
($\chi^q =$ $X$, $XX$, $X^+$ and $X^-$). The calculated alignment
of  $\chi^q$ in InAs/GaAs QDs agrees well with experimental results in
Refs. [\onlinecite{xu07b, warming09, siebert09, dalgarno08, rodt05}],
and previous calculations by Narvaez {\it et al}.\cite{narvaez05a, narvaez05b}.

\begin{figure}
\centering
\includegraphics[width=0.48\textwidth]{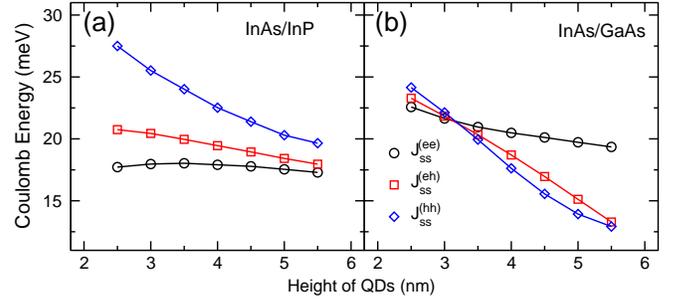}
\caption{(Color online) Diect Coulomb interactions $J_{ss}^{(hh)}$, $J_{ss}^{(eh)}$ and
  $J_{ss}^{(ee)}$ as functions of dot height
for lens-shaped (a) InAs/InP QDs and (b) InAs/GaAs QDs. The diameter of 
the QDs is fixed to 25 nm.}
\label{fig-J-K}
\end{figure}

The binding energy of $\chi^q$ is defined as,
\begin{equation}
\Delta E_{b}(\chi^{q}) = \omega(\chi^q) - \omega(X),
\label{eq-binding}
\end{equation}
where $\omega(\chi^q)$ is the corresponding recombination energy. In the Hartree-Fock 
approximation, $\omega(\chi^q)$ can be calculated as,\cite{bester03b, narvaez05b,narvaez05a}
$\omega(X) = E_g - J_{ss}^{(eh)}$,  $\omega(X^+) = E_g + J_{ss}^{(hh)} - 2J_{ss}^{(eh)}$,
$\omega(X^-) = E_g + J_{ss}^{(ee)} - 2J_{ss}^{(eh)}$,
$\omega(XX) = E_g + J_{ss}^{(ee)} + J_{ss}^{(hh)} - 3J_{ss}^{(eh)}$,
where $E_g = E_s^e - E_s^h$ (see Fig. \ref{fig-energy-diag}), $J_{ss}^{(ee)}$ ($J_{ss}^{(hh)}$)
is the direct Coulomb interaction between two electrons (holes), and $J_{ss}^{(eh)}$ is the
direct Coulomb interaction between electron and hole of $s$ levels. 
Here, we neglect the exchange interactions, which is much smaller than the
direct Coulomb interactions. With these analytical expressions, we have
\begin{eqnarray}
&& \Delta E_{b}(X^+)^{(\text{HF})} = J_{ss}^{(hh)} - J_{ss}^{(eh)}, \nonumber \\ 
&& \Delta E_{b}(X^-)^{(\text{HF})} = J_{ss}^{(ee)} - J_{ss}^{(eh)}, 
\label{eq-deltaEb}  \\
&& \Delta E_{b}(XX)^{(\text{HF})} = J_{ss}^{(hh)} + J_{ss}^{(ee)} - 2J_{ss}^{(eh)}. \nonumber
\end{eqnarray}
The values of $J_{ss}^{(hh)}$, $J_{ss}^{(eh)}$ and $J_{ss}^{(ee)}$ for
InAs/InP QDs and InAs/GaAs QDs are shown in Fig. \ref{fig-J-K}. 
For InAs/InP QDs, we see $J_{ss}^{(hh)} > J_{ss}^{(eh)} > J_{ss}^{(ee)}$. This
is because in InAs/InP QDs, holes are much more localized than electrons.
In contrast, in InAs/GaAs QDs, we find the same order for $J_{ss}^{(hh)}$,
$J_{ss}^{(eh)}$, $J_{ss}^{(ee)}$, for $h <$ 3 nm, but reversed order when $h >$ 3.0 nm.
Equation (\ref{eq-deltaEb}) can be used to explain the alignment of $\chi^q$ of InAs/InP QDs as shown in
Fig. \ref{fig-energy-pl} (a), but can not explain the alignment of $\chi^q$
in InAs/GaAs QDs. This is because in InAs/InP, $J_{ss}^{(hh)} - J_{ss}^{(eh)} \sim$ 6.2
meV, and $J_{ss}^{(eh)} - J_{ss}^{(ee)} \sim$ 3.2 meV, much larger than the correlation
energies, which is about 1 meV. However, in InAs/GaAs QDs, $J_{ss}^{(ee)}$, $J_{ss}^{(hh)}$ 
and $J_{ss}^{(eh)}$ are very close. So the correlation energies play 
very important roles to determine the final alignment of $\chi^q$. \cite{bester03b}
We therefore expect that the alignments (and binding energies) of $\chi_q$ in
InAs/InP QDs are very different from those of InAs/GaAs QDs. 

\begin{figure}
\centering
\includegraphics[width=0.48\textwidth]{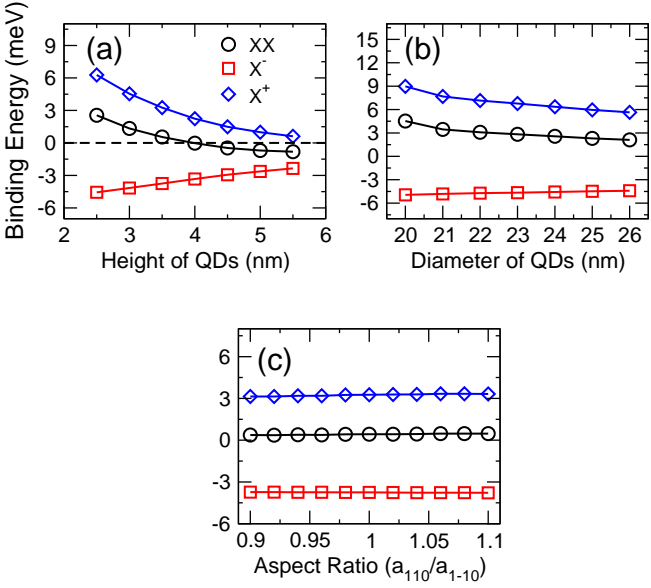}
\caption{(Color online) (a) The binding energies as functions of
dot height for lens-shaped QDs with $D$=20 nm. (b) 
The binding energies as functions of dot diameter for lens-shaped QDs
with $h$ = 3 nm, (c) The binding energies for elongated QDs, where the total volume is fixed to 
400 nm$^2$, and $h$=3.5 nm.}
\label{fig-xxxexh-InP}
\end{figure}

We show the binding energies of $X^+$, $X^-$ and $XX$ as
functions of dot height, dot diameters and aspect ratio ($a_{110}/a_{1-10}$)
 in Fig. \ref{fig-xxxexh-InP}
(a), (b) (c) respectively. 
We find that the binding energies of $X^+$ and $XX$ decrease dramatically with
the increasing of dot height, whereas the 
binding energy of $X^-$ increases.
For a lens-shaped dot with diameter $D$=20 nm, the binding energy of $XX$
equal zero, i.e., $\omega(XX) = \omega(X)$ at dot height $h \sim$ 4nm. 
At this special point, the dot can be used as an entangled photon emitter via
the time-reordering scheme proposed by Avron {\it et al}.\cite{avron08}
In contrast, the binding energy of $X^+$ is always positive, whereas
the binding energy of $X^-$ is always negative, even though they all become
small for very tall dots.
As shown in Fig. \ref{fig-xxxexh-InP} (c), there is no obvious shift of
binding energies in elongated QDs when $a_{110}/a_{1-10}$ 
changes from 0.9 to 1.1. 
The observed trends of the binding energies for InAs/InP QDs differ greatly
from those of InAs/GaAs QDs. For example, in
InAs/GaAs QDs, the binding energies of $X^+$ and $X^-$ can be either
positive or negative \cite{bester03b} due to the cross over of the Coulomb interactions 
shown in Fig. \ref{fig-J-K} . These differences can be traced
back to the difference of confinement potentials between the two dots.

\subsection{Hidden correlation}
\label{correlation}

The exciton binding energies change dramatically with the geometry and
compositions of QDs. Recently proposed by some authors of the paper 
that the hidden
correlation $\Delta$ of QDs are always positive and remarkably 
constant for a large range of self-assembled QDs 
as a consequence of the Coulomb correlation effects. \cite{zhang09}
The hidden correlation is defined as,
\begin{equation}
\Delta = \Delta E_b(XX) - \Delta E_b(X^-) - \Delta E_b(X^+),
\end{equation}
where $\Delta E_b(\chi^{q})$ has been defined in Eq. (\ref{eq-binding}) and can be extracted 
directly from Fig. \ref{fig-xxxexh-InP}. 
Under the non-self-consistent Hartree-Fock approximation,
the binding energies of $XX$, $X^+$ and $X^-$ cancel each other exactly
(see also Eq. (\ref{eq-deltaEb})), i.e., $\Delta^{\rm HF}  \equiv 0$. 
Therefore the value that $\Delta \ne 0$ arise purely from the correlation effect. 
In Ref. [\onlinecite{zhang09}], it is shown that $\Delta$ varies between 1.2 -  2.2
meV, for a wide range of geometry and exciton energies of InAs/GaAs QDs, which
agrees well with all available experimental data from different groups.\cite{zhang09} 
The hidden correlation is also confirmed by the effective
mass quantum Monte Carlo simulations.\cite{zhang09}

\begin{figure}
\centering
\includegraphics[width=0.48\textwidth]{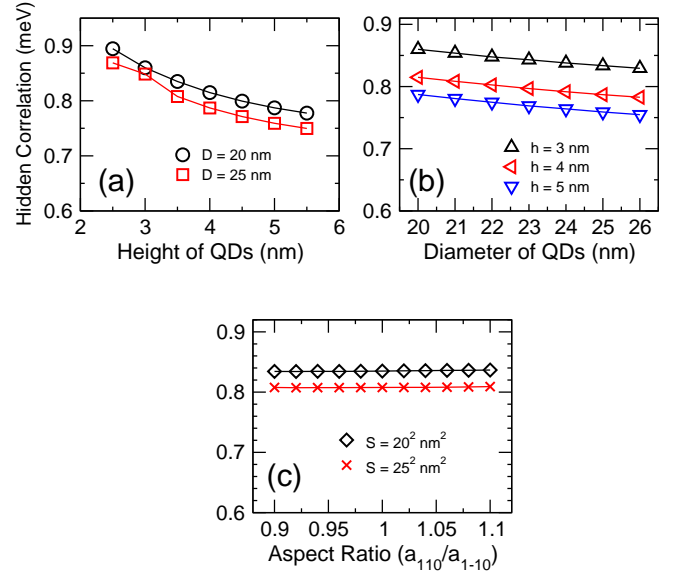}
\caption{(Color online) The hidden correlations as functions of (a) dot height. 
(b) dot diameter, and (c) aspect ratio in InAs/InP QDs. }
\label{fig-hidden-correlation}
\end{figure}

The hidden correlations in InAs/InP QDs is shown in Fig. \ref{fig-hidden-correlation},
as functions of dot heights, diameters and aspect rations.
We find $\Delta = 0.7 \sim 0.9$ meV for all QDs with different sizes and
geometries studied here.  
The hidden correlation of InAs/InP QDs is smaller than that of InAs/GaAs QDs, which can be 
understood as follows. In InAs/InP QDs, the hole is more confined, so
the energy level spacings are much larger than those in InAs/GaAs QDs (see Fig. \ref{fig-energy-diag}) and
therefore the correlation energies are reduced (see Ref. [\onlinecite{zhang09}]).
Experimental confirmation is called for this prediction.

\subsection{Lifetime of exciton, biexciton and trions}
\label{lifetime}

\begin{figure}
\centering
\includegraphics[width=0.4\textwidth]{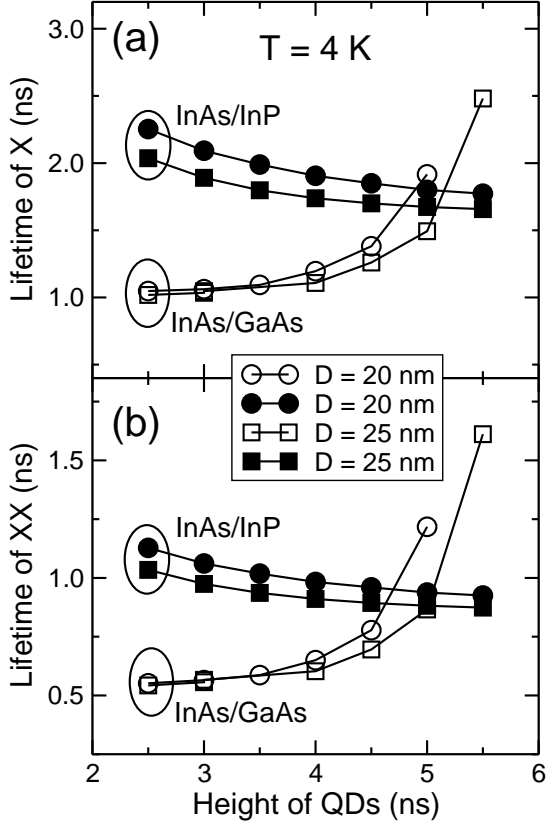}
\caption{The radiative lifetimes of (a) $X$ and (b) $XX$ in the lens-shaped InAs/InP QDs
as functions of dot height. }
\label{fig-x-xx}
\end{figure}
\begin{figure}
\centering
\includegraphics[width=0.4\textwidth]{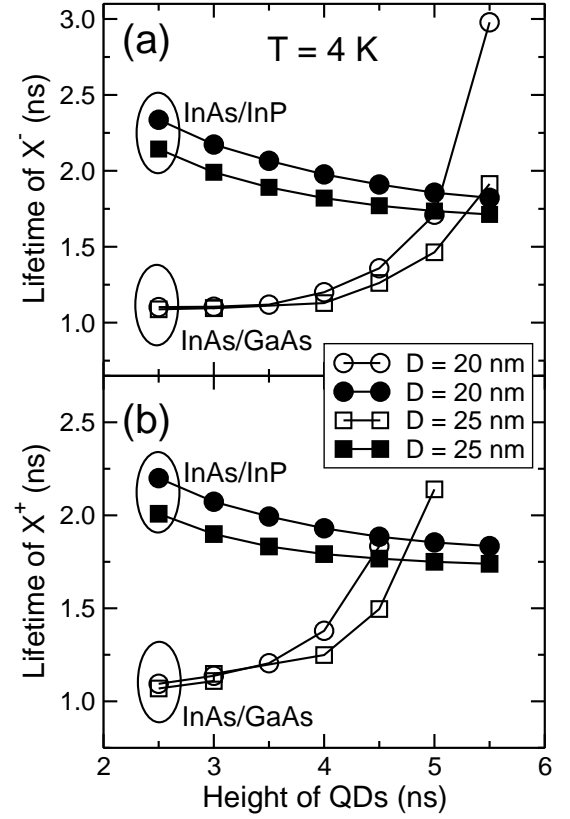}
\caption{The radiative lifetimes of (a) $X^-$  and (b) $X^+$
in the lens-shaped InAs/InP QDs as functions of dot height. }
\label{fig-xe-xh}
\end{figure}

The radiative recombination lifetimes of exciton, biexciton and trion are very important
parameters for many applications of the QDs.  At low temperature,
the lifetime is mainly determined by the dipole matrix element,
\begin{equation}
\mathcal{M}_{i, j} = \langle f|{\bf n }\cdot {\bf p} | i\rangle,
\end{equation}
where ${\bf p}$  is the dipole momentum and ${\bf n}$ is the polarization
vector of the electromagnetic field. The recombination lifetime $\tau_{if}$ at zero
temperature can be calculated from,
\begin{equation}
{1 \over \tau_{if} } =
{4e^2  n_{\text{r}} \omega_{if} \over 3m_0^2c^3\hbar^2 } |\mathcal{M}_{i, j}|^2,
\label{eq-tau}
\end{equation}
where $n_{\text{r}}$ is the reflective index, and $\omega_{if} = |\omega_i - \omega_f|$ is the recombination 
energy. $m_0$ is the mass of the electron and $c$ is the velocity
of light in vacuum.  
The linear dependence of $1/\tau_{if}$ on refractive index is applicable only when the QD 
and the matrix has similar dielectric constants. \cite{narvaez05b,
  thranhardt02}
At finite temperature, the lifetime can be calculated as,\cite{narvaez05b}
\begin{equation}
{1\over \tau(\chi^q)} = \sum_{i} { n_i \over \tau_{if}(\chi^q)}, 
\label{eq-ni}
\end{equation}
where $n_i$ is occupation number of the initio state according to the Boltzmann
distribution, and $\sum_i n_i = 1$. 
Assuming that the thermalization between the dark and bright states is much
longer than the exciton life time, we only take the bright states into
consideration, which is somehow different from the treatment in
Ref. [\onlinecite{narvaez05b}]. 

The radiative lifetimes of exciton in lens-shaped InAs/InP and InAs/GaAs QDs
are compared in Fig. \ref{fig-x-xx} (a). The major differences between the two
dots are: 
(i) The exciton lifetime in InAs/InP is almost twice longer than that
of InAs/GaAs for flat QDs. There are two reasons for the longer lifetime in
InAs/InP QDs. 
First, InAs/InP QDs have smaller exciton energies and second, 
InAs/InP QDs have much smaller dipole moment, because the electrons are less
confined.
Experimentally, the lifetimes of $X$ and $XX$ in InAs/GaAs QDs are found to be
around 0.8 - 1.2 ns, \cite{xu07a, xu07b, gammon96a, bayer02c, dalgarno08} very close
to the calculated results presented in Fig. \ref{fig-x-xx}. 
The measured exciton lifetimes in InAs/InP QDs by Sasakura 
{\it et al},\cite{sasakura09} are about 2.0 ns, which also agree
very well with the calculated values in this work. 
(ii) The exciton lifetime in InAs/InP QDs decreases with increasing of the dot
height. In the InAs/GaAs QDs, the opposite trend is found, because the hole
wave functions localize more on the interface of the dots. \cite{gong08} 
(iii) For flat QDs ($h < 3.5$ nm), we find that the exciton lifetime in
InAs/GaAs dots is almost independent of the dot diameter. However, in InAs/InP QDs, 
the exciton lifetime depends strongly on the dot diameter. For example, when $D = 20$ nm, 
$h$= 2.5 nm, $\tau(X)$= 2.26 ns. When we increase $D$ to 25 nm, $\tau(X)$ decrease to 2.04 ns.
This is also due to the weak confinement potential for electrons in the InAs/InP QDs.
Increasing the diameter of a QD results in more confined electron state, and
thus lager overlap between electron and hole wave functions and lager dipole moment.

We further compare the lifetimes of biexciton in Fig. \ref{fig-x-xx} (b), 
and of charged exciton between the two dots in Fig. \ref{fig-xe-xh}. 
The values of lifetime for these exciton complexes are very different. However
the trends of lifetimes as functions of dot geometry are rather 
similar. Roughly speaking, the charged excitons ($X^+$, $X^+$) and $X$ have lifetimes about 
twice longer than the biexciton, \cite{narvaez05b} i.e.,
\begin{equation}
\tau(X^+) \simeq \tau(X^-) \simeq \tau(X) \simeq 2\tau(XX).
\label{eq-relation}
\end{equation}

\subsection{Phase diagrams of electrons and holes}
\label{phase-diagram}

The studies of electron/hole filling process in self-assembled QDs are stimulated 
by the experiments that allow the electron/hole be loaded into the QDs one by one 
and measure the charging energies.\cite{drexler94, reuter05} 
Unlike the atoms, where large Coulomb interaction $J \sim 10 $ eV ensure
the Hund's rule (maximum spin) and Aufbau principle. In self-assembled QDs,  the Coulomb
interaction is about 20 - 25 meV for electron-electron interaction and 15 - 20 meV for
hole-hole interaction, which is the same order of the energy level spacings in QDs.
So, the electron/hole filling process should be quite different from that
in atoms. In strongly confined QDs (dimeter $\sim$ 3 nm), it is found that Hund's rule
is generally obeyed, but violation of the Aufbau principle is a common feature when
single-particle energy levels separated by a few meV.\cite{franceschetti00} In
the self-assembled InAs/GaAs QDs, He {\it et al}\cite{he06a, he05d} 
found that it is possible for holes
to violate both the Hund's Rule and the Aufbau Principe, which is confirmed by the 
PL spectra of highly charged excitons $X^{N+}$ and $X^{N-}$ by Ediger {\it et al}.\cite{ediger07a}.

Let us first look at the charging energy and addition energy of the QDs, which can
be measured  directly in the experiments.
The charging energy $\mu(N) = E(N) - E(N-1)$,  measures the energy needed to 
load an additional carrier into the QDs. The addition energy is 
the energy difference between 
$\mu(N)$ and $\mu(N-1)$, i.e., $\Delta(N-1, N) = \mu(N) - \mu(N-1) = E(N) -2E(N-1)+E(N-2)$. 
In the Hartree-Fock 
approximation, the total energy of $N$ electrons confined in 
QDs can be calculated as,\cite{franceschetti99}  
\begin{equation}
E(N) =  \sum_i \epsilon_i n_i + \sum_{i < j} (J_{i,j} - K_{i, j}) n_i n_j,
\end{equation}
where $n_i$ is the occupation number of the $i$-th level.
The diagonal Coulomb energy $J_{i,j} = \Gamma_{i,j}^{i,j}$ and the exchange energy
$K_{i,j} = \Gamma_{j,i}^{i,j}$ (see Eq. \ref{eq-gamma}).

\begin{figure}
\begin{center}
\includegraphics[width=0.48\textwidth]{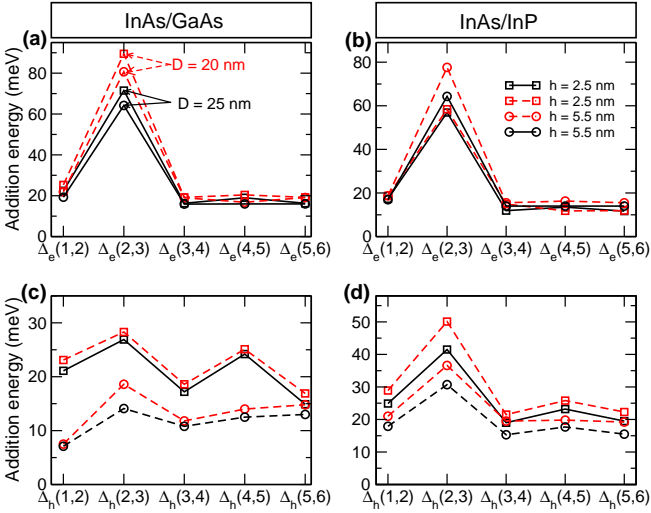}
\end{center}
\caption{(Color online) The 
electron addition energies for (a) InAs/GaAs QDs and (b) InAs/InP QDs. The 
hole addition energies are shown in (c) for InAs/GaAs QDs and in (d) for
InAs/InP QDs.
 }
\label{fig:addition}
\end{figure}

The electron/hole addition energies are presented in Fig. \ref{fig:addition}
(a),(c), for InAs/GaAs dots and in Fig. \ref{fig:addition} (b) , (d) 
for InAs/InP QDs, respectively. 
The electron addition energies of InAs/InP QDs are much smaller than
those of InAs/GaAs QDs of similar sizes. 
However, the hole addition energies of InAs/InP
QDs are much larger than those of InAs/GaAs QDs due to their much larger
single-particle level spacings, as illustrated in Fig. \ref{fig-energy-diag}.
More interestingly, where the hole addition energy in InAs/GaAs QDs show two
peaks at $N$=3 and 5, the hole addition energy in InAs/InP QDs show only a
strong peak at $N$=3, which is similar to that of electrons. The reason for
this difference is that the hole energy levels have well defined shell
structure in InAs/InP QDs, but not in InAs/GaAs QDs.
 
We further look at the filling orders of the electron/hole in the InAs/InP QDs.
The filling orders of electrons and holes in QDs are determined by the
following factors: \cite{he06a, he05d}
(i) the single-particle energy level spacing, (ii) the $p$, $d$  energy level splittings and
(iii) the Coulomb interactions. We adopt the general approaches in 
Refs. [\onlinecite{he06a, he05d}], and use the reduced 
single particle energy level spacing $\delta(p_{2},p_{1})/J_{ss}=|E_{p_2}-E_{p_1}|/J_{ss}$ and
$\delta(d_{1},p_{2})/J_{ss}=|E_{d_1}-E_{p_2}|/J_{ss}$ to construct
the phase diagrams of few carriers in QDs. 
Experimentally, the $p$-orbit splitting $\delta(p_{2},p_{1})$ in QDs
can be measured via capacitance-voltage spectroscopy \cite{reuter05} or 
terahertz pump-probe method.\cite{zibik09} 

For $N =$ 4, 5, 6, the possible ground state configurations represented 
by the spectroscopic notation read as,\cite{he05d}
\begin{eqnarray}
\label{eq-phases}
N=4:\
&& ^3\Sigma=(s^{\uparrow}s^{\downarrow})(p_{1}^{\uparrow})(p_{2}^{\uparrow}),\nonumber\\
&& ^1\Delta=(s^{\uparrow}s^{\downarrow})(p_{1}^{\uparrow}p_1^{\downarrow}); \\
N=5:\
&&^{4}\Delta=(s^{\uparrow}s^{\downarrow})(p_{1}^{\uparrow})(p_{2}^{\uparrow})(d_{1}^{\uparrow}),\nonumber\\
&&^2\Pi=(s^{\uparrow}s^{\downarrow})(p_{1}^{\uparrow}p_1^{\downarrow})(p_{2}^{\uparrow}),\nonumber\\
&&^2\Delta=(s^{\uparrow}s^{\downarrow})(p_{1}^{\uparrow}p_1^{\downarrow})(d_{1}^{\uparrow}); \\
N=6:\
&&^5\Sigma=(s^{\uparrow}s^{\downarrow})(p_{1}^{\uparrow})(p_{2}^{\uparrow})(d_{1}^{\uparrow})(d_{2}^{\uparrow}),\nonumber\\
&&^3\Pi=(s^{\uparrow}s^{\downarrow})(p_{1}^{\uparrow}p_1^{\downarrow})(p_{2}^{\uparrow)}(d_{1}^{\uparrow}),\nonumber\\
&&^1\Sigma=(s^{\uparrow}s^\downarrow)(p_{1}^\uparrow
p_1^\downarrow)(p_{2}^\uparrow p_2^\downarrow),\nonumber\\
&&^1\Sigma^*=(s^\uparrow s^\downarrow)(p_{1}^\uparrow
p_1^\downarrow)(d_{1}^\uparrow d_1^\downarrow).
\end{eqnarray}

\subsubsection{phase diagrams of electrons}

The phase diagrams of electrons in InAs/InP QDs are presented in
Fig. \ref{fig:phasediagram}(b), contrast to those from two dimensional
effective mass approximation (2D-EMA) calculations. 
The phase diagrams are determined 
by $\delta(p_2, p_1)/J_{ss}$ and $\delta(d_1, p_2)/J_{ss}$. 
Previous calculations\cite{gong08} have shown that the electron levels in
InAs/InP QDs have well defined electron structure, 
similar to those given by 2D-EMA.
Therefore the phase diagram from 2D-EMA and atomistic
model share quite similar features. The ground state configurations
for $N$ = 4 - 6 are $^3\Sigma$, $^2\Pi$, $^1\Sigma$ (indicated by the red
spots in the figure), respectively, following 
Hund's rule and Aufbau principle. However, for $N$=4, the area allow
stable $^3\Sigma$ configuration
in InAs/InP QDs is much smaller than that of the 2D-EMA model.

\subsubsection{phase diagrams of holes}

\begin{figure}
\begin{center}
\includegraphics[width=0.48\textwidth]{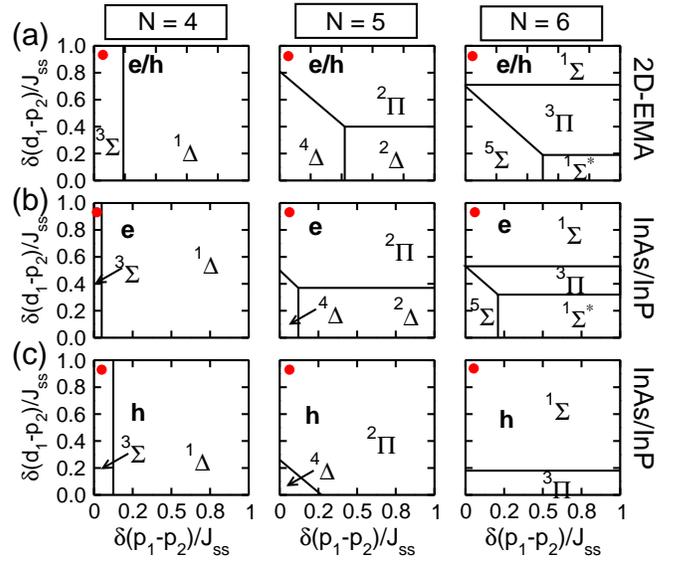}
\end{center}
\caption{Phase diagrams of InAs/InP QDs with number of electrons/holes $N$ = 4, 5, 6.
(a) The results from 2D-EMA calculations for both electrons and holes, which are identical for 
the parabolic confinement potential. (b) The results of InAs/InP QDs from
atomistic pseudopotential method for electrons and (c) for holes. The red dots
indicate the configurations of the ground state.}
\label{fig:phasediagram}
\end{figure}

The phase diagrams of holes in InAs/InP QDs are presented in
Fig. \ref{fig:phasediagram}(c).  Unlike those of the InAs/GaAs QDs, the hole levels 
in InAs/InP QDs have well defined shell structure. The ground state of $N = 3$
is $^3\Sigma$, satisfying Hund's rule. 
However, the phase diagrams of holes in the 
InAs/InP QDs still differ greatly than calculated by 2D-EMA for $N$ = 5, 6. 

{\it $N = 5$}: The are three possible ground state configurations for 2D-EMA model and in 
InAs/GaAs QDs.\cite{he05a}  He {\it et al}\cite{he05a} have predicated that the non-Aufbau 
phase $^2\Delta$ can be found for large $\delta(p_2, p_1)$ in InAs/GaAs QDs. Similar phase,
however, can not be found in InAs/InP QDs. 
This can be understood as following. In InAs/InP QDs, the $p$-orbital splitting
$\delta(p_{2},p_{1})\sim$ 1 - 4 meV and $p$-$d$ level spacing $\delta(d_1, p_2) \sim$ 15 - 30 meV
(see Fig. \ref{fig-energy-diag}), so the Coulomb interactions between holes are not strong enough 
to overcome the $p$-$d$ spacing in InAs/InP QDs. Therefore $^2\Delta$ is
always unfavorable in energy in InAs/InP dots. 

{\it $N = 6$}: There are four possible ground state configurations
for $N = 6$. In both InAs/GaAs and InAs/InP QDs, the $^5\Delta$ can not be observed. 
However, there are still obvious difference between the filling orders in InAs/GaAs and 
InAs/InP QDs.  In InAs/GaAs QDs,\cite{he05a} the hole can be localized in the interface, 
resulting in extremely small level spacing between $p_2$ and $d_1$ orbits, hence the  $^1\Sigma^*$,
which violate the Hund's rule, is possible.\cite{he06a, he05d, ediger07a} 
In InAs/InP QDs, only two phases, $^1\Sigma$ and $^3\Pi$, can be found.

The hole energy level spacings in InAs/InP QDs are significantly larger than in
InAs/GaAs dots, and the $p$ level splittings are significantly
smaller. Therefore, the
ground states for $N$=5, 6 in InAs/InP QDs are $^2\Pi$ and $^1\Sigma$
respectively, as indicated by the red dots in Fig.\ref{fig-energy-diag}(c),
which obey both Hund's 
rule and Aufbau principle.

\subsection{PL Spectra of highly charged excitons}
\label{multi-x}

Having clarified the ground state configurations and filling orders of
electrons and holes in the InAs/InP QDs, we now turn to the PL spectra of
highly charged excitons, which provides a useful tool to explore the
complex interactions in the QDs.\cite{ediger07a}
In InAs/GaAs QDs, it is found that the PL spectra of
highly charged excitons exhibit some peculiar 
properties due to the breakdown of the Aufbau principle for hole.\cite{ediger07a}

\begin{figure}
\centering
\includegraphics[width=0.48\textwidth]{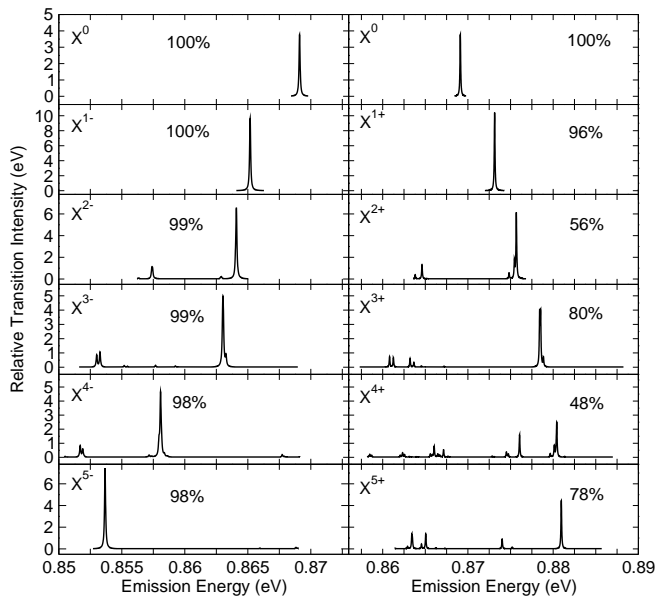}
\caption{The PL spectra of highly charged excitons in a lens-shaped
InAs/InP QD with $D$=25 nm and $h$ = 2.5 nm at 4 K. 
The number in each panel is the pecentage of the leading configuration
of the initial state CI wave functions.}
\label{fig-multi-x-InP}
\end{figure}

We present the calculated PL spectra of $X^{N+}$ and $X^{N-}$ ($N$ = 0 - 5)
for InAs/InP QDs in Fig. \ref{fig-multi-x-InP}. We also calculated the PL spectra of highly
charged InAs/GaAs QDs. The results (not shown) agree very well with previous calculations 
in Ref. [\onlinecite{ediger07a}]. There are several differences between the PL
spectra of these InAs/InP and InAs/GaAs dots. 

(i) The transition intensities of $X^{N+}$ and $X^{N-}$ in 
InAs/InP are weaker than that in InAs/GaAs QDs, due to the relatively longer lifetime 
in InAs/InP QDs. 
Rough estimation shows that the transition intensities in InAs/InP are 
half of those in InAs/GaAs QDs.

(ii) In InAs/InP QDs, the main peak of $X^{5-}$ is on the lower energy side of  $X^0$ by
about 16 meV, and that of $X^{5+}$ is on the higher energy side of $X^0$ by
about 12 meV, much larger than the energy shifts in InAs/GaAs QDs, 
where both $X^{5-}$ and $X^{5+}$ are 
on the lower energy side of  $X^0$ by about 10 meV and 2 meV respectively. 
These differences are due to the different confinement potentials of the two
dots, as discussed in Sec. \ref{bindingenergy}.

(iii) The number in each panel is the overlap between the many-particle wave functions
from CI calculations and those from Hartree-Fock approximations for the
initial states, i.e.,
\begin{equation}
\mathcal{O}(X^{\alpha}) =  |\langle \Psi_{\rm CI}(X^{\alpha})|\Psi_{\rm
  HF}(X^{\alpha})\rangle|^2\, .
\end{equation}
$\mathcal{O}(X^{\alpha})$ represents the percentage of the leading configurations in state $X^{\alpha}$.
We find that there are significant differences of $\mathcal{O}(X^{\alpha})$ between
$X^{N-}$ and $X^{N+}$ in InAs/InP QDs. For $X^{N-}$, 
$\mathcal{O}$ is generally larger than 98\%, in contrast to 95\% in InAs/GaAs QDs. 
Moreover, the configuration hybridization of the final states of
 $X^{N-}$ ($N$ = 4, 5) is also much smaller than that in the InAs/GaAs QDs. 
Therefore,  $X^{4-}$ and  $X^{5-}$ have much fewer peaks in the InAs/InP 
PL spectra than in the InAs/GaAs spectra\cite{ediger07a}.
For $X^{N+}$, we find that the hybridization effect in InAs/InP is much stronger
than that in InAs/GaAs QDs. For example, in InAs/InP QDs, $\mathcal{O}(X^{2+})$ = 56\%,
much smaller than 84\% in InAs/GaAs QDs. This giant difference is from the smaller $p$ 
level energy splitting in InAs/InP QDs, as illustrated in  Fig. \ref{fig-energy-diag}. The
strong hybridization effects lead to the much more complex $X^{N+}$ 
PL spectra in InAs/InP QDs when $N \ge$ 3. 

(iv) In the InAs/GaAs QDs, holes may violate Hund's rule and Afubau's
  principle. For example, in the InAs/GaAs QDs, 
 $X^{3+}$ has a close shell ground state,\cite{ediger07a} and therefore no polarization
  dependent photoluminescence. In contrast, $X^{3+}$ of InAs/InP QDs 
  has a open shell ground state, and therefore has polarization
  dependent photoluminescence. Furthermore, there are many more transition
  peaks due to the spin splitting in the InAs/InP spectra than in the
  InAs/GaAs spectra.
  Previous calculations \cite{ediger07a} show that $X^{5+}$ has open shell
  ground state and has polarization dependent photoluminescence in the
  InAs/GaAs QDs. However, in the InAs/InP QDs, $X^{5+}$ has close shell
  ground state, and no polarization
  dependence in the photoluminescence. These differences reflect that the hole
occupation in the two dots are significantly different.

\section{Summary}
\label{conclusion}

In this work, we present a comprehensive studies of the optical properties of InAs/InP
QDs using an empirical pseudopotential method and configuration interaction
treatment of many-particle interactions. The results are contrast to those of 
InAs/GaAs QDs, whose properties are well understood, both theoretically and
experimentally. 
The main difference between the optical properties of the two QDs can be
summarized as:
(i) The alignment of emission lines of neutral exciton, charged exciton and biexciton
in InAs/InP QDs is quite different from that in InAs/GaAs QDs.
(ii) The hidden correlation in InAs/InP QDs is about 0.7 - 0.9 meV, much smaller than that
in InAs/GaAs QDs (1.2 - 2.2 meV).
(iii) The lifetimes of neutral exciton, charged exciton and biexciton in InAs/InP QDs
are about twice longer than those in InAs/GaAs QDs. 
(iv) The phase diagrams of electrons and holes in InAs/InP QDs differ greatly from those in
InAs/GaAs QDs. The filling orders of electrons and holes are shown to obey the Hund's rule
and Aufbau principle.
(vi) The PL spectra of highly charged excitons show some significant
difference between the two dots, due to the different filling orders and Coulomb
interactions between carriers.

\acknowledgments
L.H. acknowledges support from the National Science 
Foundation of China under grants 60921091
and ``Hundreds of Talents'' program from Chinese Academy of Sciences.


\end{document}